\RequirePackage{fix-cm}
\documentclass[smallextended]{svjour3}       % onecolumn (second format)
\smartqed  % flush right qed marks, e.g. at end of proof
\usepackage{graphicx}
\usepackage{pgfplots}

\pgfplotsset{compat=1.11,
    /pgfplots/ybar legend/.style={
    /pgfplots/legend image code/.code={%
       \draw[##1,/tikz/.cd,yshift=-0.25em]
        (0cm,0cm) rectangle (3pt,0.8em);},
   },
}

\begin{document}

\title{IMHOTEP - Virtual Reality Framework\\for Surgical Applications
%\thanks{Grants or other notes
%about the article that should go on the front page should be
%placed here. General acknowledgments should be placed at the end of the article.}
}
%\subtitle{Do you have a subtitle?\\ If so, write it here}

%\titlerunning{Short form of title}        % if too long for running head

\author{Micha Pfeiffer\footnote{These authors contributed equally to this work.}         \and
        Hannes Kenngott\footnotemark[1]			\and
        Anas Preukschas			\and
        Matthias Huber			\and
        Lisa Bettscheider		\and
        Beat M\"uller-Stich		\and
        Stefanie Speidel
}

\authorrunning{Micha Pfeiffer et al.} % if too long for running head

\institute{Micha Pfeiffer, Stefanie Speidel \at
              National Center for Tumor Diseases, Dresden, Germany \\
%              Tel.: +123-45-678910\\
%              Fax: +123-45-678910\\
              \email{micha.pfeiffer@nct-dresden.de}           %  \\
%             \emph{Present address:} of F. Author  %  if needed
%           \and
%           S. Author \at
%              second address
			\and
			Matthias Huber			\at
			Karlsruhe Institute of Technology, Institute for Anthropomatics and Robotics
			\and
			Hannes Kenngott, Anas Preukschas, Lisa Bettscheider, Beat M\"uller-Stich		\at
			Heidelberg University Hospital, Department of General-, Visceral- and Transplant Surgery
}
	
\date{Pre-print version.  Accepted: March 2018. \\Published in "International Journal of Computer Assisted Radiology and Surgery"\\Doi: 10.1007/s11548-018-1730-x}
% The correct dates will be entered by the editor

\maketitle

\begin{abstract}

\emph{Purpose} The data which is available to surgeons before, during and after surgery is steadily increasing in quantity as well as diversity.
When planning a patient's treatment, this large amount of information can be difficult to interpret.
%In clinical routine, the large amount of information can hinder the physicians' ability to interpret it.
To aid in processing the information, new methods need to be found to present multi-modal patient data, ideally combining textual, imagery, temporal and 3D data in a holistic and context-aware system.
\\
\emph{Methods}
We present an open-source framework which allows handling of patient data in a virtual reality (VR) environment. By using VR technology, the workspace available to the surgeon is maximized and 3D patient data is rendered in stereo, which increases depth perception. The framework organizes the data into workspaces and contains tools which allow users to control, manipulate and enhance the data.
%Due to the framework's modular design, it can be extended by user-defined tools and can be used as a research platform in clinical visualization, simulation and patient data organization. \\
Due to the framework's modular design, it can easily be adapted and extended for various clinical applications. \\
\emph{Results}
The framework was evaluated by clinical personnel (77 participants). The majority of the group stated that a complex surgical situation is easier to comprehend by using the framework, and that it is very well suited for education.
Furthermore, the application to various clinical scenarios - including the simulation of excitation-propagation in the human atrium - demonstrated the framework's adaptability.
As a feasibility study, the framework was used during the planning phase of the surgical removal of a large central carcinoma from a patient's liver.
%The framework's extendability was tested in student research projects, including tools for bone segmentation and the simulation of excitation-propagation in the human atrium.
\\
\emph{Conclusion}
The clinical evaluation showed a large potential and high acceptance for the VR environment in a medical context. The various applications confirmed that the framework is easily extended and can be used in real-time simulation as well as for the manipulation of complex anatomical structures.

% \PACS{PACS code1 \and PACS code2 \and more}
% \subclass{MSC code1 \and MSC code2 \and more}
\end{abstract}

\section{Introduction}

\label{intro}

The amount of data recorded before, during and after a surgery has drastically increased in recent years, due to the availability of new imaging technologies as well as better equipment for many non-imaging modalities. Supplementary information such as segmentations and annotations can be generated and data from other surgeries and patients can be used for comparison. Ideally, all of these information sources should be considered when planning the treatment of a patient.
Managing the expanding amount of information will be increasingly difficult in the foreseeable future, and displaying the data to surgeons in a meaningful context will require new visualization and rendering systems.

\begin{figure*}
  \includegraphics[width=\textwidth]{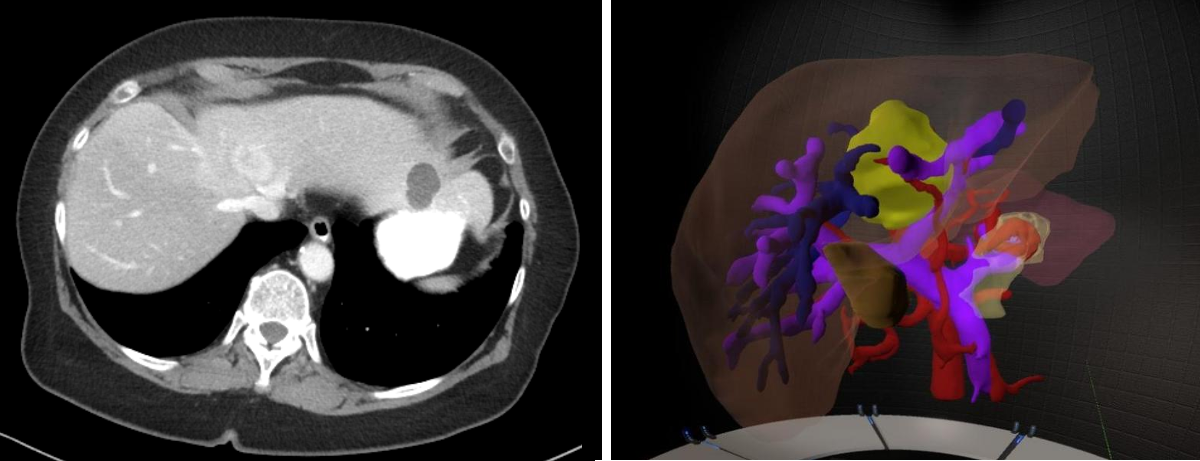}
	\caption{Left: Traditionally, 2D slices of CT scans are used for planning. Right: In complex scenarios such as this one, where the tumor (yellow) and the various vessel trees (veins, arteries and bile ducts) lie close together, a 3D visualization could greatly benefit physicians in analyzing patient data (screenshot of the virtual reality application).}
	\label{fig:Liver}       % Give a unique label
\end{figure*}

An important portion of the acquired data is inherently three-dimensional, such as CT and MRI multi-slice scans. Usually, this data is viewed as individual slices on conventional 2D screens, resulting in the loss of depth perception, even though it has been shown that viewing data in 3D greatly enhances the comprehension of the presented structures \cite{Herfarth2002,Mueller_Stich2013}. When planning the treatment of a patient, however, it is vital that the surgeons understand the precise spatial location of tumors, organs and vessels to optimize a possible surgical procedure and minimize patient risk (Fig.~\ref{fig:Liver}).

We present the IMHOTEP\footnote{Immersive Medical Hands-On Operation Teaching and Planning System} framework, which meets the need to visualize very inhomogeneous data types in an intuitive, simple to navigate virtual reality (VR) environment. The framework provides tools for loading, viewing and manipulating patient-specific data. By rendering the scene in stereo (one image for each eye, with slightly shifted perspective between the images), depth perception is increased. We use Head-Mounted-Displays (HMDs) to expand the available workspace to a full $360^\circ$ virtual environment.
By using modern gaming technology we achieve frame rates high enough for real-time interaction.

The aim of the framework is to be a development and research platform for VR tools in clinical scenarios. To this end, we provide many standard functions which facilitate the use of a VR environment in conjunction with medical data. The core functionalities of the framework are open-source and are available online\footnote{https://github.com/IMHOTEP-Medical/imhotep}. The framework's central design choices are discussed below, while implementation details and instruction for usage can be found in the project's documentation. Readers are encouraged to experiment with the project and add custom tools to develop their own medical VR applications.

%\begin{figure*}
%  \includegraphics[width=0.8\textwidth]{User2.png}
%	\caption{Virtual reality headset worn by surgeon to view patient data.}
%	\label{fig:User}       % Give a unique label
%\end{figure*}

\subsection{Areas of Application}

\emph{Pre-Operative Planning:}
Complex surgical operations require individual planning on a per-patient basis. Many factors have to be considered, such as patient age, allergies and medication taken by the patient, as well as the patient history and previous surgeries. By visualizing the data in VR, the surgeons are able to quickly view relations in the data within one single application and to review the anatomical situation, even before performing the first incision.

\emph{Simulation:}
Understanding complex dynamic 3D scenes is demanding. By providing the tools to port such animations into VR, the framework can be used to display simulations in stereo 3D, allowing users to view the animation almost as if they were present in real life. The framework's intuitive input tools can be used to manipulate the simulation settings and the effects can be viewed in real-time.

\emph{Teaching and Learning in Virtual Reality:}
Current teaching material usually only shows projections of anatomical structures, which can make them difficult to understand. Simple learning tasks can be integrated into the framework quickly. By expanding the framework with animations and simulations, it can aid in the understanding not only of 3D but also of 4D (3D plus time) processes.

\subsection{Related Work}

The potential of VR in the field of data visualization has been widely recognized \cite{WebVR2017,Donalek2014,Moran2015}\footnote{While older publications often use the term "Virtual Reality" to refer to a very wide range of methods, we use it only for those systems in which the user wears a Head-Mounted-Display, with stereo rendering and positional as well as orientational tracking.}. VR data visualization systems have been created for the areas of climate research \cite{DICE2016}, social media feed data \cite{Perhac2017} and software analysis \cite{Vincur2017}. Companies such as \emph{Kineviz}\footnote{https://www.kineviz.com/}, \emph{Virtualitics}\footnote{https://www.virtualitics.com/} and \emph{Dynamoid}\footnote{http://dynamoid.com} already offer VR data visualization applications for businesses, science and engineering. They use data mining algorithms and machine learning with the goal of making big data easier to analyze.

In the medical domain, the usage of VR applications has quickly increased in recent years \cite{Khor2016}, mainly due to the development of low-cost hardware. Many of these applications are focused on treating patients with traumas \cite{Valmaggia2016,MaplesKeller2017,Atherton2016} or chronic pain \cite{Schmitt2011}. These applications place patients into an immersive virtual scene, often as a form of exposure therapy.

In contrast, VR applications for physicians often focus on medical education. These range from visualizations of anatomical structures \cite{He2017} and tools for surgical training \cite{Huber2017,Mathur2015} to virtual scenarios which aim to enhance the interaction between clinical personnel and patients \cite{Goldman2016,Zielke2017}. In another project, 3D video captured during surgeries is played back and experienced in VR \cite{Cha2016}.
%Recently,  YOU (biolucid.com)
These systems are usually used with a fixed data set (such as a generic patient model or pre-recorded video) and do not support the examination of patient-specific data for clinical use. They are not focused on clinical day-to-day routine and usually rely on older data sets.

Only recently, the first VR systems have been developed which are designed for diagnostic purposes. Examples are tools for the analysis of pathology slides \cite{Pathology2016} and tools which visualize the changes in sizes and shapes of tumors over time in a 3D virtual setting \cite{Reddivari2017}. Ard et al. \cite{Neuro2017} have developed a VR system which visualizes brain MRI scans as well as several augmentations of the data.  While these projects show the significance of new VR tools in medicine, they are aimed only at single areas of application and lack a broader, flexible and reusable architecture.

Egger et al. \cite{Egger2017} implemented a VR plug-in for the medical image processing software \emph{MeVisLab}. The plug-in focuses on the rendering of the virtual scene and relies on MeVisLab for loading and management of the data. However, this method only moves small portions of the available data into VR, so relationships between the inhomogeneous data sources cannot be visualized and context awareness may be lost.

Using our framework, Loewe et al. \cite{CinCVR} have created an interactive VR ablation simulation which allows physicians to explore the effects which ablation lesions have on atrial arrhythmias. Their work focuses on the usage of the framework for one specific application. In contrast, this article describes the underlying framework itself, its system components and design as well as a more general evaluation.

\section{Methods}

VR-based surgical assistance tools have several requirements. They need to
\begin{itemize}
\item handle common clinical file formats and protocols,
\item smartly position the inhomogeneous data types,
\item render the virtual environment in stereo to allow for true 3D vision,
\item feature an intuitive user interface tailored for clinical day-to-day usage,
\item handle interaction with the VR hardware
\item ensure a high frame rate to create an enjoyable VR experience.
\end{itemize}
Since these conditions are independent of the exact application, a platform which abstracts away the required functionality can considerably speed up the development of VR tools for clinical usage.
Our framework combines techniques from modern gaming technology with medical viewers. It provides functions to control the asynchronous loading, visualization and organization of patient data and handles user input and interaction with the hardware.
%organization of patient data, user input and graphical user interface as well as interaction with the hardware.
%organization of patient data and also handles user input and interaction with the hardware

\begin{figure*}
  \includegraphics[width=\textwidth]{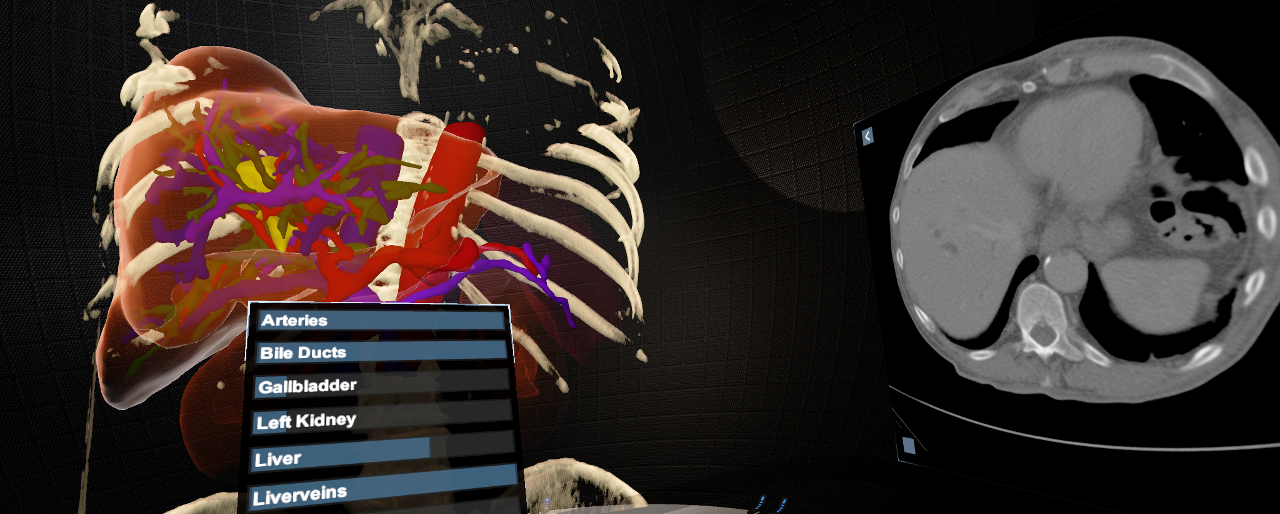}
\caption{The virtual scene. The patient has been configured to show the surface rendering of the liver, liver vessels and tumor (yellow), along with a volumetric rendering of the rib cage. CT slice images are displayed on the virtual 2D screen for reference. On the bottom left, part of a hand-held tool palette can be seen, displaying a tool to control organ transparency.}
\label{fig:VR_Overview}       % Give a unique label
\end{figure*}

\begin{figure*}
  \includegraphics[width=\textwidth]{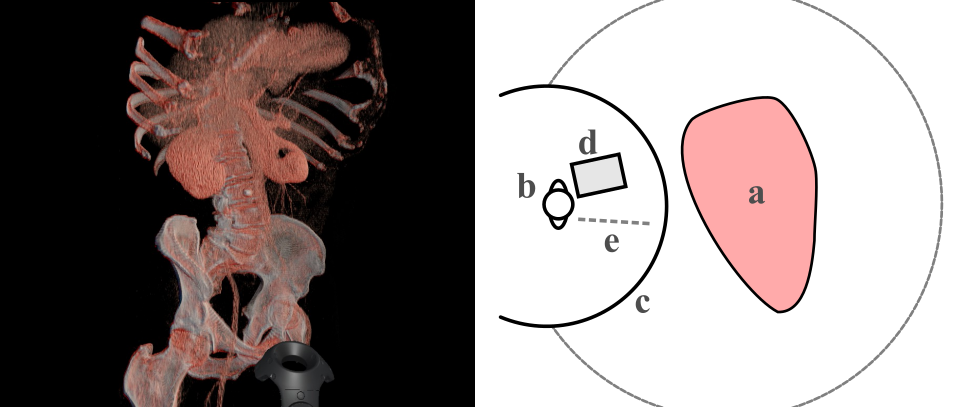}
\caption{Left: Volumetric rendering of a CT scan of a patient's abdomen. The chosen transfer function extracts the skeleton and some organs, but can be modified to display different parts of the volume. Right: Schematic top view of the workspace layout. The 3D data is displayed in the center of the virtual room (a). The user (b) stands on a platform on the side of the room, surrounded by a large curved transparent 2D screen (c). The user has a tool palette (d) which follows them around. A virtual laser pointer (e) is used to interact with the UI and the data.}
\label{fig:LayoutAndVolumetric}
\end{figure*}

\subsection{Patient Data - From Disk to Visualization}
At the core of the framework lies the loading and visualization of patient-specific data sets. These data sets can be comprised of textual information, 2D images, 3D volumes and 3D surface meshes (Fig.~\ref{fig:VR_Overview}).

For the visualization of the 3D data, the framework provides two methods:
If 3D volumes (such as the slices resulting from a CT or MRI scan) are available, they can be displayed via volumetric rendering (Fig.~\ref{fig:LayoutAndVolumetric}, left).
%This is realized by a custom sliced volume rendering method which runs on the GPU.
This is realized by a custom GPU shader program which renders 3D textures using a slice-based approach \cite{VolumeRendering1994} and a Blinn-Phong lighting model \cite{Blinn1977}.
Transfer functions assign colors to the individual voxels in the data. By changing the transfer functions, different parts of the data can be visualized and highlighted. The rendering process is done twice for each time frame, once for each eye, generating stereo vision. The volumetric data is loaded in the common DICOM\footnote{Digital Imaging and Communications in Medicine} standard via the ITK\footnote{Insight Segmentation and Registration Toolkit} library.

If previously segmented surface meshes exist, these can be displayed in the virtual scene (Fig.~\ref{fig:Liver}, right). The meshes (representing structures such as organs, vessels and tumors) are generated using external tools. They are loaded into the \emph{Blender3D} software - which can import many common file formats - and are then saved in Blender's native \emph{.blend} file format. Individual colors can be assigned to the various meshes before they are opened and displayed by the framework.
% To display organs and structures, surface meshes are loaded. These are often the result of a segmentation process and a subsequent method to generate a triangulated surface mesh (such as the Marching Cubes algorithm\cite{cn}) run on CT or MRI data, but other sources are possible. To support multiple file formats, the meshes are imported in the open-source \emph{Blender3D} software and each organ or structure can be assigned a color before being saved as a Blender file, which can then be loaded in the framework.
% The 2D DICOM images (or slices of the 3D volumes) can also be displayed. Users can scroll through Stacks-of-Slices as usually generated by CT or MRI.
%If required, these images can also be displayed in the 3D workspace and registered to the surface meshes, to enhance the perception of position.

Additional data (such as laboratory results and patient history) can be loaded and displayed alongside the image data. Optionally, text can be formatted in HTML to allow text coloring, font formatting and individual text layouts depending on the application.

%\begin{figure*}
%  \includegraphics[width=\textwidth]{VR_Volumetric.png}
%\caption{Stereo rendering of volumetric data. The chosen transfer function displays the patient's bones and kidneys. Choosing a different transfer function shows different parts of the data without the need for a segmentation.}
%\label{fig:VR_Volumetric}       % Give a unique label
%\end{figure*}

%\begin{figure*}
%  \includegraphics[width=\textwidth]{VR_LiverSingle.png}
%\caption{Renderings of the surface mesh of a patient's liver inside the VR environment. Fully opaque rendering (left) shows the structural layout and shape of the organs. By changing the organ transparency, users can view the internal structures, such as the tumor and vessels (right).}
%\label{fig:VR_Liver}       % Give a unique label
%\end{figure*}

\subsection{Data Organization: Introducing Workspaces}
The type of information which is accumulated before, during and after a surgery is extremely diverse. First, there is data which is best presented as text, such as patient data (name, age, sex, diagnosis). Next, there is data which is inherently 2D, such as images (slices from MR- and CT-scans, photographs or images from endoscopic video), graphs (development of patient data over time) and flow charts. Finally, there is 3D data such as the surface meshes of the patient's organs, along with annotations. Additionally, most of the tools for data manipulation need space in which to display their input elements and current status. In order to make best use of the available space and to present each data type in an intuitive way, the virtual scene is divided into three distinct workspaces (Fig.~\ref{fig:LayoutAndVolumetric}, right).

%\begin{figure*}
%\begin{tabular}{cc}
%\includegraphics[width=0.5\textwidth]{WorkspacesLayout.eps} &
%\includegraphics[width=0.45\textwidth]{VR_OpacityControl.png} \\
%\end{tabular}
%\caption{table caption...}
%\end{figure*}

\emph{3D Workspace:} To fully exploit the benefits of the 3D display techniques available in VR, the largest workspace in the center of the spherical virtual room is dedicated to displaying the 3D patient data. This automatically draws the initial attention of the user to the 3D visualization of the patient's organs. Since the space is so large, the organs can be scaled to a larger-than-life size, allowing the viewer to focus on small details.

\emph{2D Screens:} To supplement the 3D data with additional 2D information, a curved screen is introduced, which surrounds the viewer in the virtual scene. The resulting area is much larger than that of conventional 2D screens and the user can view the entire screen simply by turning their head. Positions of the various types of data (such as images or patient meta data) can be pre-defined so that similar data will always be found on a similar position on the screen.

\emph{Tool Palette:}
Most of the available tools have settings which can be manipulated by the user. These settings can be accessed by choosing the tool from the \emph{tool ring}, a user interface element fixed to one of the input controllers. Once a tool has been chosen, its options are displayed in the \emph{tool palette} next to the controller. In this way, active tools are always easy to access while inactive tools do not use up any of the available screen space.

\subsection{Data Control Tools}
The intention of the framework is to be used in very different applications. To implement these, researchers can add their own tools which build on the framework's interface. Some commonly used standard tools are already present in the framework:

\emph{Annotation Control:}
Annotations are a simple way of marking structures and positions in 3D space and are commonly used in 3D design and visualization. The framework contains an annotation tool which can be used to highlight important positions, areas and volumes. Each marker can be linked to a textual label to add additional information. To minimize overlap between the labels and the patient data, \emph{hedgehog labeling} \cite{Hedgehog2014} is used to automatically place the labels.

%\emph{Opacity Control:}
%In a lot of surgical situations, the structures of interest, such as vessels or tumors, are hidden by tissue and other organs. To help with the studying of internal structures, the opacity control tool can be used to make individual surface meshes transparent or temporarily hide them entirely (see also Fig. \ref{fig:VR_Liver}).

\emph{View Control:}
Surgeons and radiologists are used to viewing data from certain orientations, such as for example the sagittal, transverse and coronal views. The view control tool can be used to pre-define such views so that the data from different patients will be rotated and aligned to standard views automatically.

\subsection{User Interaction}
The HTC Vive's tracked controllers are used as the main input system. While other methods, such as tracking the user's hands or gestures, might seem more natural at first, they were less suited for productive work, since precise selection and positioning of objects was not possible. By linking their movement to actions like rotating the patient organs or zooming the 2D DICOM images, user interaction becomes very intuitive. To precisely point at positions and select objects, a virtual laser pointer tool is attached to one controller.
%There are many different input methods available today, including inexpensive sensors which track the user's hands (\emph{Leap Motion}\footnote{https://www.leapmotion.com/}), gesture-control sensors (\emph{Myo Armband}\footnote{https://www.myo.com/}). However, user studies quickly showed that both methods were not well suited for productive work. Selecting and pointing at precise positions was impossible with the Myo and cumbersome with the Leap Motion. Data gloves may be an alternative, but they are usually more expensive, and are less suited for interacting with objects which are further away than an arm's length. Instead, the system integrates the HTC Vive tracked controllers. By linking their movement to actions like rotating the patient organs or zooming the 2D DICOM images, user interaction becomes very intuitive. To precisely point at positions and select objects, a virtual laser pointer tool is attached to one controller.

\subsection{Implementation Details}

\begin{figure*}
  \includegraphics[width=\textwidth]{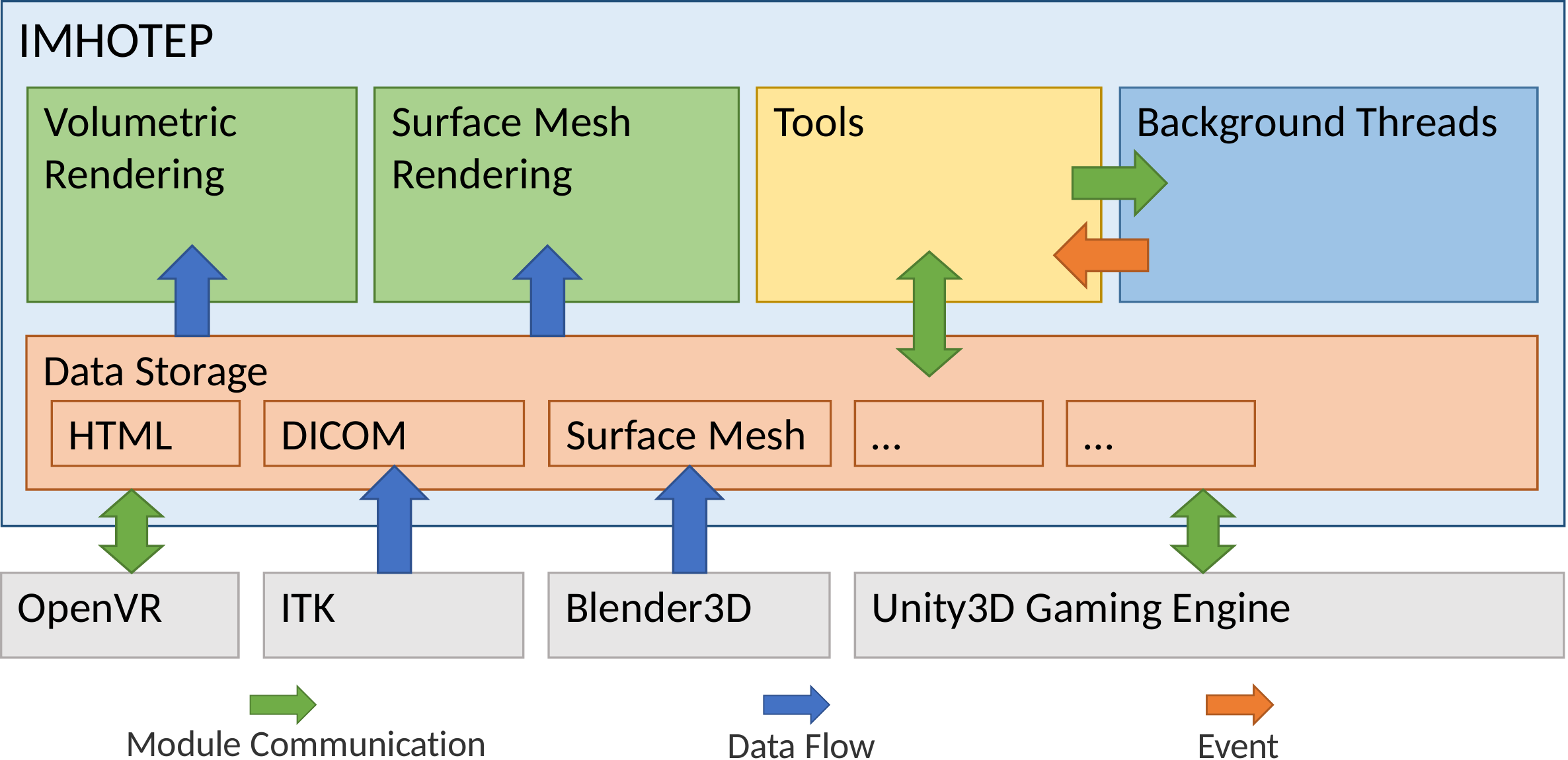}
%\caption{IMHOTEP architecture overview. The framework builds on the Unity3D engine. It uses ITK to parse DICOM data, reads Blender3D surface meshes and communicates with VR hardware through OpenVR\protect\footnotemark. The various data types can be manipulated by tools and visualized using various methods. Demanding tasks can be outsourced to separate threads. Asynchronous tasks can communicate across modules by firing events.}
  \caption{IMHOTEP architecture overview. The framework builds on the Unity3D engine. It uses ITK to parse DICOM data, reads Blender3D surface meshes and communicates with VR hardware through OpenVR. The various data types can be manipulated by tools and visualized using various methods. Demanding tasks can be outsourced to separate threads. Asynchronous tasks can communicate across modules by firing events. Tested with Unity3D Version 2017.2, SimpleITK Version 1.0.1, OpenVR/SteamVR Version 1.2.1 and Blender3D Version 2.78. }
\label{fig:SoftwareStructure}       % Give a unique label
\end{figure*}

In order to achieve real-time performance, the framework uses the Unity3D engine as the rendering back-end. The engine is well suited for the task since it is aimed at game development (an area where interactive frame rates are necessary). To ensure extendability, the framework's architecture is composed of individual modules which exchange information through application programming interfaces and threads (Fig.~\ref{fig:SoftwareStructure}).

\emph{Supported Hardware and Modes:}
The system supports both the \emph{Occulus Rift} as well as the \emph{HTC Vive} and auto-detects which of the two HMDs is available. Depending on the supported functions of the detected hardware, the framework switches between \emph{Room-Scale-VR} where the user can walk around the scene and \emph{seated} mode. Furthermore, a development mode allows to use the program without any headset.

\emph{Events and Threads:}
IMHOTEP builds upon an event system which limits module inter-dependencies and realizes extendability. If one module fires an event, an arbitrary number of other modules can be notified and react individually. The available core events can be extended with other, tool-specific events. Since essential functions in a medical visualization framework require a lot of processing time, such as the loading of large data sets or calculation of a simulation response, expensive tasks are outsourced to background threads. By using the threading system in conjunction with the event system, tasks can be run asynchronously and then fire an event to notify other modules upon completion.

\section{Results}

%\subsection{Surgical Operation Planning}

% for example if tumor is deep inside a vital organ or close to important vessels
% the relation of tissue which is to be removed to that which is to be preserved,

\begin{figure*}
\begin{tikzpicture}
\begin{axis}[
%	x tick label style={
%			/pgf/number format/1000 sep=},
	xtick=data,
	xticklabels={1 (Very low), 2, 3, 4, 5 (Very high)},
	yticklabels={,,},
	%ylabel=Testers,
	xtick pos=left,
	ytick pos=left,
	ytick=\empty,
	enlargelimits=0.15,
	legend style={at={(0.5,-0.15)},
		anchor=north west,legend columns=1,
		legend pos=north west},
	ybar=2pt,% configures `bar shift'
	bar width=11pt,
%	ylabel near ticks,
	nodes near coords,
	%point meta=y *10^-7 % the displayed number
	width=\textwidth,
	height=0.35\textwidth,
	enlarge y limits={0.25,upper},
	legend cell align={left}
]
\addplot
	coordinates {(1,0) (2,2) (3,10) (4,21) (5,44)};
\addplot
	coordinates {(1,0) (2,1) (3,14) (4,29) (5,33)};
%\addplot
%	coordinates {(1,8) (2,14) (3,27) (4,34) (5,29)};
\legend{Student education, Postgraduate education}
	\end{axis}
\end{tikzpicture}
\caption{Participants' responses to the questions: How high do you estimate the potential of this technology for a) student education ($\mu=4.39$, $\sigma=0.81$) and b) postgraduate education for physicians ($\mu=4.22$, $\sigma=0.79$)?
Given values represent the mean $\mu$ and standard deviation $\sigma$ of the responses.}
\label{fig:Education}       % Give a unique label
\end{figure*}
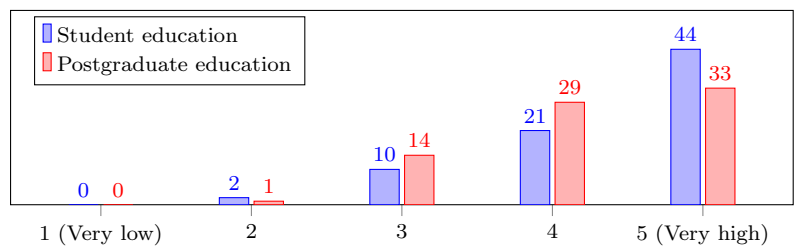

\emph{User Study:} Using a patient data set for a complex hepatectomy, a user study was performed with 77 test persons (25 resident physicians, 30 medical students and 22 clinical staff). After viewing the data in VR, the test persons were asked to assess the application in terms of clinical and educational usefulness. Most participants stated that the application is beneficial for the planning of complex surgery, both in terms of quality as well as speed at which the anatomical situation can be understood (Fig.~\ref{fig:ResultsComplexCase}). The participants also believed the tool to be useful for the education of medical students and postgraduates (Fig.~\ref{fig:Education}). Of the 77 participants, most stated that the experience in the virtual scene was comfortable (37) or very comfortable (33).

\emph{Operation Planning:} To evaluate the feasibility in a clinical preoperative planning setting, a pilot study was conducted using the data of a 75 year old female patient with a large central hepatocellular carcinoma comprising the segments I, IVa, V and VIII. The segmented 3D surface models were displayed semi-transparently to allow for the visualization of all target and risk structures without the need to change the view. CT images, laboratory parameters and  the patient diagnoses and current medication were loaded into the system. Before surgery the surgeons revisited the patient information first in the classical way in the hospital information system (HIS) and Picture Archiving and Communication System (PACS). Their judgment of the case was based on the information in these systems. In addition they used the system to visualize the patient's data and records in VR.

The surgeons who evaluated the system confirmed qualitatively that it may lead to a faster and possibly better preoperative reassessment of relevant clinical data. They believed the system was able to show all relevant information of the patient in the VR environment at point of care. The data could be generated in a reasonable time frame considering the pilot study character of the evaluation. Visualizations could be handled easily to virtually generate the planned direct line-of-sight of the surgeons during operation.

\begin{figure*}
\begin{tikzpicture}
\begin{axis}[
%	x tick label style={
%			/pgf/number format/1000 sep=},
	xtick=data,
	xticklabels={1 (Not at all),2,3,4,5 (Very much)},
	yticklabels={,,},
	%ylabel=Testers,
	xtick pos=left,
	ytick pos=left,
	ytick=\empty,
	enlargelimits=0.15,
	legend style={at={(0.5,-0.15)},
		anchor=north west,legend columns=1,
		legend pos=north west},
	ybar=2pt,% configures `bar shift'
	bar width=11pt,
	ylabel near ticks,
	nodes near coords,
	%point meta=y *10^-7 % the displayed number
	width=\textwidth,
	height=0.35\textwidth,
	enlarge y limits={0.25,upper},
]
\addplot
	coordinates {(1,0) (2,2) (3,11) (4,32) (5,32)};
\addplot
	coordinates {(1,1) (2,3) (3,9) (4,39) (5,25)};
\legend{Better, Faster}
	\end{axis}
\end{tikzpicture}
\caption{Participants' responses to the questions: By using the tool, can a complex surgical situation be understood a) better ($\mu=4.22$, $\sigma=0.79$) and b) faster ($\mu=4.09$, $\sigma=0.85$) when compared to traditional software?
Given values represent the mean $\mu$ and standard deviation $\sigma$ of the responses.}
\label{fig:ResultsComplexCase}       % Give a unique label
\end{figure*}
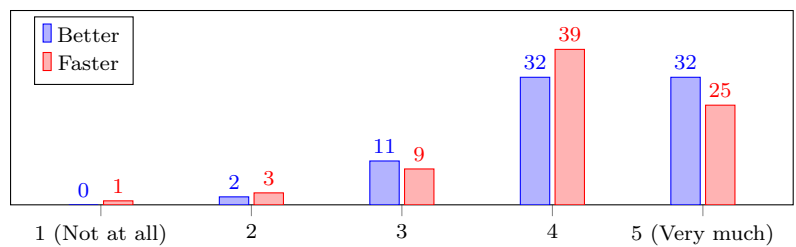

\begin{figure*}
  \includegraphics[width=\textwidth]{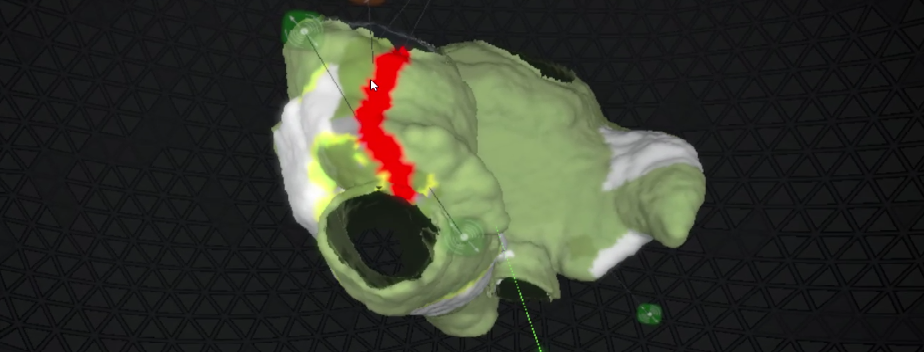}
\caption{Screenshot of the ablation simulation of the human atrium \cite{CinCVR}. The color of the surface mesh is animated to show the activation wave (yellow), the refractory regions (green) and regions which have not yet been activated (white). The user can draw ablation lesions (red) and study their influence on the wave's propagation.}
\label{fig:Atrium}       % Give a unique label
\end{figure*}

\emph{Extendability:} The aforementioned real-time capable ablation simulation system \cite{CinCVR} was built on top of our framework. The authors added their own fast-marching algorithm to simulate excitation propagation, as well as tools which allow the user to add and remove ablation lesions, handle tissue types, manipulate pacemakers and control simulation time. By relying on our framework's ability to work with patient-specific data, their aim is to provide a tool to study the effect of lesions on arrhythmic behaviors in the human atrium for diagnosis as well as therapy planning (Fig.~\ref{fig:Atrium}).
Their user study showed that the application has a short learning curve due to the intuitive user interface and leads to a greater understanding of arrhythmias.

%Another extension to the framework was made to study the possibility to perform medical segmentation inside a VR environment \cite{Traechtler2017}. This work used the framework's user interface to build intuitive tools for the advanced manipulation of 2D as well as 3D data. It relied on the framework's threading interface to perform demanding segmentation tasks without interfering with the rendering process.

%In another work, the framework was used to build an application for liver resection planning \cite{Wallisch2016}. Functions were added to allow the positioning of resection planes, to perform the cut on the liver surface meshes and to calculate residual liver volumes. The implementation of this tool benefits not only from the framework's user interface and threading functions but also from the mesh manipulation functions provided by the underlying Unity3D engine.

\section{Discussion}

The clinical evaluation of the framework has shown a very high acceptance for the use of VR in the context of surgical applications amongst medical personnel and students. Most users found the interaction with the data very intuitive and natural.

The framework has successfully been used to build an operation planning tool as well as an ablation simulation tool. Various other projects are being developed on top of the framework, including a resection planning system to allow the placing of resection planes and the calculation of residual volume.
These results are encouraging when considering the future of VR in medical applications.
%sEven so, some open questions still need to be addressed:

The studies performed here were conducted using qualitative questionnaires and can only summarize subjective opinions of the test persons. To objectively confirm the benefit of using a VR system when viewing complex data, further studies have to be carried out. Ideally, these should evaluate criteria such as cognitive load on the users, their ability to interpret a complex anatomical situation, time needed for operation planning and operation outcome and should be compared to results from a control group.

\

The framework can be used as-is to realize many medical applications. For more advanced usages, additional functionalities would be needed:

\emph{Muli-User-Environment:}
For complex cases, a multi-user extension of the framework would allow doctors to view and discuss the treatment with experts from all over the world in a virtual tumor board. In this area, open questions are the real-time interaction with the visualizations by multiple users at once, secure transfer of data and automatic anonymization.

\emph{Educational Uses:}
Participants in the user study expect a high benefit from using the VR environment for educational purposes. To aid in the development of such tools, functions common to various educational applications should be implemented in the framework such as quiz and scoring systems, the rendering of realistic tissue textures and gaming elements.

\emph{Clinical Integration:}
To make the system easier to use in a clinical day-to-day scenario, it should be integrated into the hospital information system (HIS), to allow rendering patient data directly from hospital servers. This is currently limited since required data - such as surface meshes - is usually not present for most patients, but would make the usage of VR in hospitals much more feasible in the near future.

%\begin{itemize}
%	\item [x] Tumor-Board over network
%	\item [x] Segmentation
%	\item Realistic Textures, sick vs healthy tissue
%	\item Soft-Tissue Simulation
%	\item [x] Cutting
%\end{itemize}

\begin{acknowledgements}
%If you'd like to thank anyone, place your comments here
%and remove the percent signs.
We thank the \emph{Medien- und Filmgesellschaft Baden-W\"urttemberg} for supporting early development of the framework through the \emph{Karl-Steinbuch} scholarship.
\end{acknowledgements}

\section{Compliance with Ethical Standard}
\noindent \textbf{Conflict of Interest:} The authors declare that they have no conflict of interest.

%\noindent \textbf{Protection of Human Rights} For the feasibility study, ethics approval has been granted by the local ethics committee (Ethics Committee of the Heidelberg Medical Faculty).
\noindent \textbf{Ethical Approval:} All procedures performed in studies involving human participants were in accordance with the ethical standards of the institutional and/or national research committee and with the 1964 Helsinki declaration and its later amendments or comparable ethical standards.

\noindent \textbf{Informed Consent:} Informed consent was obtained from the study participant.
%\noindent \textbf{Ethical approval} All procedures performed in studies involving human participants were in accordance with the ethical standards of the institutional and/or national research committee and with the 1964 Helsinki declaration and its later amendments or comparable ethical standards.

% BibTeX users please use one of
%\bibliographystyle{spbasic}      % basic style, author-year citations
\bibliographystyle{spmpsci}      % mathematics and physical sciences
%\bibliographystyle{spphys}       % APS-like style for physics
%\bibliography{}   % name your BibTeX data base
\bibliography{IMHOTEPLiteratureShort}

\end{document}